\documentstyle[eqsecnum,aps,preprint]{revtex}
\begin{document}
\draft
\title{Realistic interactions and dilepton production off pp-collisions}
\draft
\author{F. de Jong and U. Mosel}
\address{Institut f\"ur Theoretische Physik, Universit\"at Giessen,
35392 Giessen, Germany}
\date{\today}
\maketitle
\begin{abstract}
We present a model for dilepton production of proton-proton collisions
using a realist $T$-matrix that by incorporating $\Delta$ degrees of
freedom fits the $NN$ scattering data up to 2 GeV. 
The results we find differ in details from earlier work that use less
sophisticated interactions but the overall agreement with these
calculations is good.
\end{abstract}


\section{Introduction}

Recently various scenarios related to in-medium modifications of the 
properties of the $\rho$-meson have been
proposed \cite{Chanfray,Li} to explain the apparent increase in the 
production of dilepton pairs with an invariant mass around the $\rho$ meson 
in heavy ion collisions \cite{Ceres}.
To really be able to extract these signals, possibly due to a decrease in the
$\rho$ mass as a result of chiral symmetry restoration, the description of
the fundamental processes has to be on a firm ground and its
inherent approximations have to be well understood. 
In this paper we will investigate the elementary process of dilepton pair 
production off nucleon-nucleon collisions. 
Two ingredients are of importance here. 
First there is the question of the nucleon-nucleon form factor.
Vector Meson Dominance suggests a form which naturally gives a significant
increase in the cross-section around the $\rho$ mass. 
However, while the concept is well established for the pion formfactor, 
for the nucleon formfactor its validity is much less clear \cite{Doenges}.
In this paper we will focus on a second point, the off-shell character of 
the $NN$ interaction. 
From real photon bremsstrahlung one knows that already at pion threshold 
off-shell effects are crucial to describe the data.
For virtual photon bremsstrahlung one might expect the effects to be
even larger since the typical energies involved are much higher
than for real photon bremsstrahlung and one thus can reach kinematics
which are further off shell. 
In this paper we will present a calculation based on a $T$ matrix
which includes $\Delta$ degrees of freedom and fits the $NN$ data
up to 2 GeV. 
We will compare our results with a previous calculation that uses
a One Boson Exchange fit to the $NN$ on-shell data to find its off-shell 
behaviour \cite{Schaefer}.  

\section{Theoretical Framework}

The quantity of interest is the differential cross-section for production of 
lepton pairs with an invariant mass $M$.
It is most easily calculated in the c.m. frame of the 
incoming nucleon pair
\begin{eqnarray}
\frac{d \sigma}{d M} &=& \frac{1}{p_{cm} \sqrt{s}} 
\frac{e^4 m_N^4 m_e^2}{8 (2 \pi)^7 M^4}
\int_0^{P_\gamma^{max}} dP_\gamma \frac{M P^2_\gamma}{P^0_\gamma}
d \Omega_\gamma \nonumber \\
&\times& \int d \Omega_{p_N} 
\frac{2 p^2_N}{-P_\gamma cos(\theta_{p_N}) + 2 p_N(E^N_{+} + E^N_{-})}
M_\mu M_\nu
\int d \Omega_{p_e} 
\frac{2 p^2_e}{P_\gamma cos(\theta_{p_e}) + 2 p_e(E^e_{+} + E^e_{-})}
L^{\mu \nu} \nonumber \\
\mbox{with} & & P_\gamma^{max} = [ \frac{1}{16s}
\left(2s - 8m^2 - M^2\right) - M^2 ]^\frac{1}{2}.
\end{eqnarray}
In this expression $e$ is the elementary charge, $m_N,m_e$ the nucleon and
electron masses respectively, $s$ the total incoming invariant
mass and $p_{cm}$ the momentum of the incoming nuclei.
Also,
$p_N$ is the momentum of the outgoing nucleons
relative to each other, its value is determined by energy
conservation.
The final nucleon pair thus has momenta 
$p_N^{+} = -\frac{P_\gamma}{2} + p_N, p_N^{-} = -\frac{P_\gamma}{2} - p_N$,
the energies of the final nucleon pair are denoted by $E^{+}_N, E^{-}_N$.
In the integration over the solid angle of $p_N$, $\theta_{p_N}$ is taken 
relative to the photon momentum, $\bar{P_\gamma}$.
A similar notation is used for the dilepton pair, their momenta are
$p_e^{+} = \frac{P_\gamma}{2} + p_e, p_e^{-} = -\frac{P_\gamma}{2} - p_e$,
with corresponding energies $E^{+}_e, E^{-}_e$.
Furthermore, $P^0_\gamma$ is the zero-th component of the photon momentum,
$P^0_\gamma = \sqrt{P_\gamma^2 + M^2}$. 
$L^{\mu \nu}$ is the leptonic tensor describing the decay of the virtual
photon into the dilepton pair,
\begin{equation}
L^{\mu \nu} = \frac{1}{m_e^2} [p^\mu_{e,+}p^\nu_{e,-} + 
p^\mu_{e,-}p^\nu_{e,+} - g^{\mu \nu} (p_{+} \cdot p_{-} + m^2_e)].
\end{equation}
$M^\mu$ is the matrix element for producing a virtual photon with 
rest mass $M$ off a nucleon-nucleon collision. 
The diagrams we include in the calculation of this amplitude are shown 
in Fig. \ref{diagrams}. 
We have the nucleon single-scattering (\ref{diagrams}a), rescattering
(\ref{diagrams}b) and $\Delta$ single-scattering (\ref{diagrams}c) and
rescatter diagrams (\ref{diagrams}d). 
We include the same diagrams as in a previous study of real-photon 
bremsstrahlung \cite{FdJ_ppg2}, the expressions given there are readily 
generalized to the case of virtual photon production.

For the $NN\gamma$ vertex we again use the standard form without formfactors
\begin{equation}
\Gamma^{NN\gamma}_\mu = -ie \gamma_\mu - 
e \kappa \frac{\sigma_{\mu\nu} k^\nu}{2m_N},
\end{equation}
where we defined the photon momentum $k = p_i - p_f$ to be outgoing.
The form of the $N \Delta \gamma$ vertices is less well established. 
Following Jones and Scadron \cite{Jones} we have the general form
\begin{eqnarray}
\Gamma_{\mu\nu}^{N \Delta \gamma} &=&
K^1_{\mu\nu} + K^2_{\mu\nu}  + K^3_{\mu\nu} \nonumber \\
\Gamma_{\mu\nu}^{\Delta N \gamma} &=&
-K^1_{\mu\nu} + K^2_{\mu\nu}  + K^3_{\mu\nu} \hspace{5mm} {\rm with} \nonumber \\
K^1_{\mu\nu} &=& ie G_1 (g_{\mu\nu} \mbox{$\not \! k$} -  k_\mu \gamma_\nu) \gamma^5 T_z \nonumber \\
K^2_{\mu\nu} &=& ie G_2 (g_{\mu\nu} P \cdot k -  k_\mu P_\nu) \gamma^5 T_z \nonumber \\
K^3_{\mu\nu} &=& ie G_3 (g_{\mu\nu} k^2 - k_\mu k_\nu) \gamma^5 T_z.
\end{eqnarray}
In these expressions the index $\mu$ is to be contracted with an index of the 
$\Delta$ propagator,
$P = p_\mu^\Delta + p_\mu^N$ is the total momentum and
$T_z$ is third component of the isospin transition matrix for coupling
an isospin 3/2 to an isospin 1/2 particle.
The $K^3$ component does not contribute in the case of a real photon ($k^2 = 0$) and thus
experimental input on this quantity is very limited. 
One attempt is the calculation of Nozawa and Lee \cite{Nozawa_II} of electroproduction
of pions off the nucleon.
However, using the values Nozawa and Lee employ we found that results were 
insensitive to the contribution from this vertex part and we can safely 
ignore this contribution.

The most elaborate determinations of the $G_1$ and $G_2$ coupling constants
come from fitting the $M1^+$ and $E1^+$ multipole data on the 
photoproduction of pions off nucleons 
\cite{Jones,Nozawa,Davidson,Blomqvist,Koch,Garcilazo}.
In judging these numbers we have to realize that in our calculation  
we treat the nucleon background differently. 
It is therefore the most realistic to take values from calculations 
where the nucleon background contribution in the 
$P_{33}$ channel is implicitly included in the coupling constants. 
In such an approach Jones and Scadron find  
$G_1 = 2.68$ (GeV$^{-1}$) and $G_2 = -1.84$ (GeV$^{-2}$), which we
will use in our calculation. 
This is very close to the values found in a recent analysis of 
Lee \cite{Lee_4}:
$G_1 = 2.89$ (GeV$^{-1}$) and $G_2 = -2.18$ (GeV$^{-2}$).
With these numbers we have to bear in mind that the dominant contribution of 
the vertex comes from the $K_1$ part and we can order the sets of values 
according to the value of $G_1$, 
the accompanying value of $G_2$ has not much influence on the results.
An independent method to find the values is assuming vector-dominance. Then the 
values are determined by the ratio $g_{NN\rho}/g_{N\Delta\rho}$. 
Using the values of ter Haar \cite{terHaar} we find
$G_1 = 2.0$ (GeV$^{-1}$) and $G_2 = 0.0$ (GeV$^{-2}$).
This value is on the low end of the range of values found in
pion-photoproduction.
In Ref. \cite{Schaefer} a value taken from the decay width of the $\Delta$ 
into a photon and nucleon is used: 
$G_1 = 2.3$ (GeV$^{-1}$) and $G_2 = 0.0$ (GeV$^{-2}$).

In calculating the diagrams we use a $T$-matrix that is based on the Paris
potential and additionally contain $\Delta$-degrees of freedom. 
Including $\Delta$ degrees of freedom is essential for describing the 
inelastic channels and resonant structures in the phase-shifts.
The resulting $T$-matrix describes the nucleon-nucleon data well up 
to 2 GeV \cite{Lee_1}.

Other calculations of dilepton production use a less sophisticated 
description of the $NN$ interaction.
One approach is to develop a Soft Photon Approximation for virtual photon 
bremsstrahlung as is done in Ref. \cite{Korchin,Zhang}.
Another possibility is to parametrize the on-shell nucleon scattering
data by means of an One Boson Exchange interaction, and use this 
parametrization to extrapolate to the off-shell matrix elements which
enter in the calculation of the matrix element $M^\mu$ 
\cite{Schaefer,Brat,Haglin}.
This method has certain limitations, first of all, the NN-scattering
amplitude is not unitary due to its OBE form. 
Also, in the fit to the $NN$ data in the approach of Ref. \cite{Schaefer}
the on-shell equivalence of the pseudo-scalar and pseudo-vector coupling
makes the parameters independent of whether one chooses a 
pseudo-scalar or pseudo-vector pion-nucleon-nucleon vertex.
However, the results for dilepton production do depend on this choice.
Although the contribution from the electric part of the vertex remains the 
same \cite{Schaefer_2},
the contribution from the magnetic part of the vertex is affected
by this choice. 
Using a $T$-matrix one has a more realistic description of the off-shell
character of the $NN$ interaction, since the off-shell amplitudes enter in 
the T-matrix equation and these are thus implicitly constrained by the fit to
the $NN$ data. 
For example, the degeneracy of the pseudo-scalar/pseudo-vector choice for the 
pion coupling is resolved to a large extent: one has to use different 
parameters to obtain the same fit.

Using a sophisticated description of the $NN$ interaction has 
ramifications on other parts of the model.
The current arising from the nucleonic diagrams 
(Figs. \ref{diagrams}a,b) is not gauge-invariant since we 
do not include negative energy states in the intermediate propagators
of the $T$-matrix.
Note that the $\Delta$-decay diagrams are gauge-invariant due to the
gauge invariant $N \Delta \gamma$ vertex. 
For real photon bremsstrahlung this is not too big a problem. 
In that case the current is gauge invariant up to order $k$ in the 
photon momentum. 
Moreover, an actual calculation of the negative energy state contributions
shows them to be small \cite{FdJ_ppg3}.

Due to the fact that the matrix element for virtual photon bremsstrahlung
depends on two independent quantities (the mass of the virtual photon and
its momentum) the situation for virtual photon-bremsstrahlung is less
favourable. 
Only in the c.m. frame of the virtual photon one can show that the matrix
element is model independent only up to order $\frac{1}{M_\gamma}$.
One also finds rather large differences between the various low
energy theorems \cite{Korchin}. 
This all implies that the issue of gauge invariance is more important for
dilepton pair production than it is for real photon bremsstrahlung.  

The problem of a non-gauge invariant current is also encountered in 
models of $(e, e'p)$ reactions. 
Various schemes have been devised to render a calculated current 
gauge invariant \cite{Forest,Naus,Caballero,Mougey}.
Of course neither scheme is unique and will lead to different 
results which can be related to the choice of a particular 
gauge \cite{Pollock}.
Having noticed that our $M_0$ is particulary large as compared to the 
spatial components we decided to use the method of 
Refs. \cite{Naus,Caballero} and express $M_0$ in terms of $M_i$ to obtain 
a conserved current:
\begin{equation}
M_0 = \frac{P^i_\gamma M_i}{P^0_\gamma}.
\end{equation}
The resulting amplitude then trivially fullfils the gauge-invariance
condition. 
This choice has the advantage that in this scheme the results
for real-photon bremsstrahlung are unaffected since this process is
independent of $M_0$.

\section{Results}

With the model described before we calculated the differential cross-section
for production of dilepton pairs with invariant mass M for proton-proton 
collisions at 1 GeV laboratory energy. 
We restrict ourselves to proton-proton collisions since in neutron-proton 
collisions meson-exchange-currents usually dominate, but are very elaborate
to calculate in a $T$-matrix approach. 
We think that studying the pp reaction provides a clearer picture of the 
additional dynamics introduced by the use of a $T$-matrix. 
We choose a laboratory energy of 1 GeV to still be in the range where the
$T$-matrix provides a detailed description of the phase-shifts. 
At higher energies the total and inelastic cross-sections are still very well 
reproduced, but the reproduction of more sensitive observables like 
the polarization cross-sections is less satisfactory \cite{Lee_1}.
Since the latter is also true for other models that fit the $NN$ data
to high energies using a $T$-matrix formalism \cite{Machleidt}, 
one might argue that the description of 
$NN$ scattering in terms of nucleon and meson degrees of freedom in
terms of a $T$-matrix starts to break down and one has to include additional
physics, like explicit quark degrees of freedom.

The results are displayed in Fig. \ref{results_1}.
Comparing our nucleonic contribution (dash-dot-dot) with the one 
of Sch\"afer {\it et al.} (dash-dot) we see a rather large difference.
At low $M$ our cross-section is around a factor 2 larger than the one
of Sch\"afer. At higher $M$ the situation is reversed.
There the cross-section of Sch\"afer is larger. 
This can be largely attributed to the choice of a pseudo-scalar pion
vertex. 
Taking a pseudo-vector pion coupling the result of Sch\"afer is reduced by
a factor 10 and is again smaller than our result.
The Soft-Photon result of Ref. \cite{Zhang} (short-dashed line) is even lower 
than the  Sch\"afer result, and we can conclude that in our calculation the 
additional off-shell information we include leads to an enhancement of
the nucleonic part of the cross-section.
At the energies under consideration the total cross-section is dominated by 
the $\Delta$ contributions, even more
than we found for the real-photon bremsstrahlung case \cite{FdJ_ppg2}.
To show the dependence of the total results on the strength of the $N\Delta\gamma$
vertex we performed calculations with a strong coupling 
$G_1 = 2.68$ (GeV$^{-1}$) and $G_2 = -1.84$ (GeV$^{-2}$, dotted line)
and with the relatively weak vector dominance value
$G_1 = 2.0$ (GeV$^{-1}$) and $G_2 = 0.0$ (GeV$^{-2}$, dashed line).
The result of Sch\"afer
$G_1 = 2.3$ (GeV$^{-1}$) and $G_2 = 0.0$ (GeV$^{-2}$) is represented by the
solid line. 
The result with the weak coupling is about 25 \% smaller than the one with the
strong coupling. 
As stated before, these values are on the high and low end of the spectrum
found in various pion-photoproduction models and thus represent a natural
measure of the theoretical uncertainty.
The results of Sch\"afer {\it et al.} nicely fall in between and
the overall agreement between both models is reassuring. 
It remains to be seen whether the similarities hold when looking at more
refined observables, like e.g. anisotropies of the dilepton pairs
\cite{Brat}.
Another point of additional investigation is the influence of form-factors.
A straightforward application of vector dominance will give us simply
an upward shift in the cross-section, increasing with increasing
invariant mass $M$.
However, a more detailed calculation of the form-factors, as performed
in Ref. \cite{Doenges} shows that the vector dominance assumption is too
general for the nucleonic form-factor. 

In conclusion, we presented a calculation of dilepton production at 1 GeV
laboratory energy using a $T$ matrix that includes $\Delta$ degrees of
freedom and provides an excellent fit to the $NN$ scattering data. 
We argued that the use of a $T$ matrix gives a more reliable description of
the off-shell behaviour of the effective $NN$ interaction, the price to 
be paid however is the non-gauge invariance of the calculated current. 
We showed that after repairing gauge invariance in one of the well-known 
schemes, we find results that are similar to the ones obtained in models
that use less sophisticated $NN$ interactions.
In particular, we verified that the dominant contribution to virtual
bremsstrahlung in this energy range comes from the $\Delta$-resonance
decay. 
In detail we found significant differences: the nucleonic contribution
is larger than found in models which only include on-shell $NN$ scattering
information.

The authors would like to thank prof. T.-S. H. Lee for making his $T$-matrix
available to us and valuable comments on the manuscript. This work was 
supported by GSI Darmstadt and BMBF.

\begin{figure}
\caption{The various diagrams included in the current.}
\label{diagrams}
\end{figure}

\begin{figure}
\caption{
Differential cross-section for dilepton pair production off pp-collisions
at 1 GeV laboratory energy as a function of the invariant mass $M$. 
The solid line and dash-dot line are the
full and nucleonic result of Sch\"afer {\it et al.}
The dotted line represents our full result calculated with the strong
$N\Delta\gamma$ coupling constant, the long-dashed line the one with the
weak coupling constant. The dash-dot-dot line stands for our nucleonic
result. 
The short-dashed line is the Soft-Photon result of Eq. 8
of \protect\cite{Zhang}.
}
\label{results_1}
\end{figure}

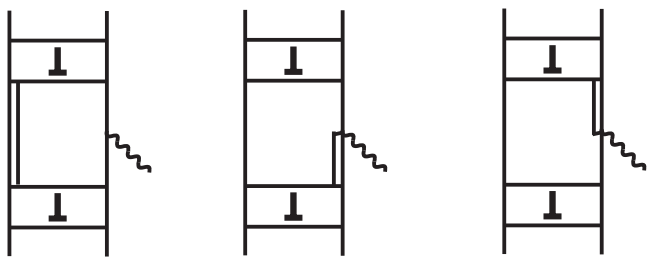
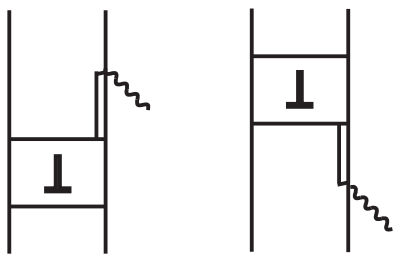
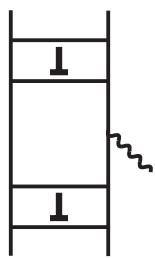
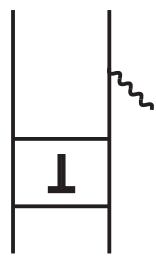
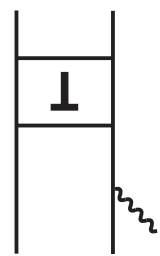
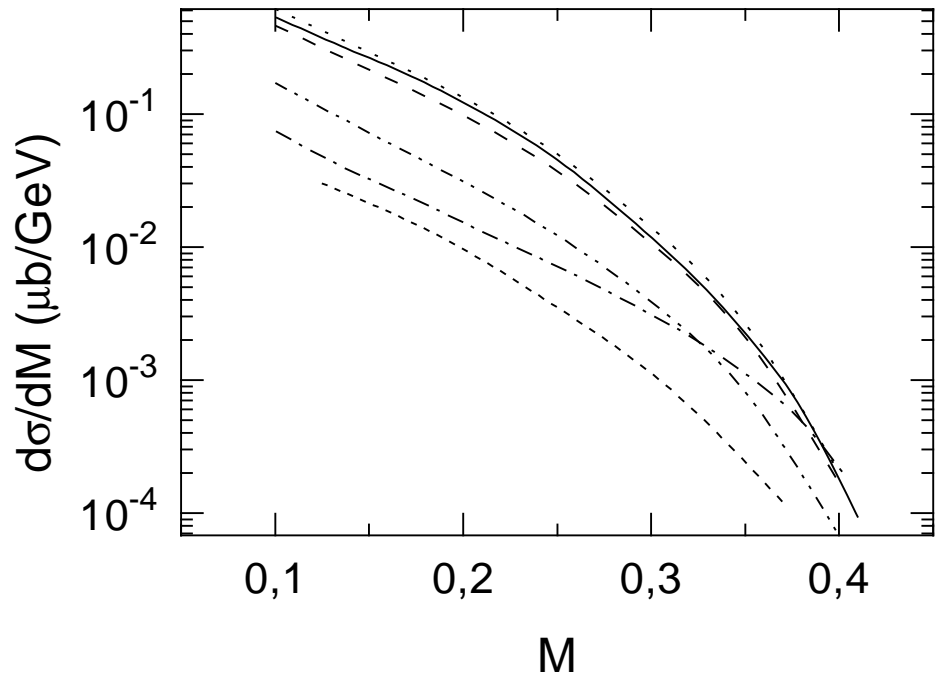
\end{document}